\documentclass[preprint,showpacs,preprintnumbers,amsmath,amssymb,prb,aps]{revtex4-1}
\usepackage{epsfig}
\usepackage{epstopdf}
\usepackage{graphicx}
\usepackage{dcolumn}
\usepackage{bm}
\usepackage{slashbox}
\usepackage{textcomp}
\begin{document}

\title{Role of orbital degrees of freedom in investigating the magnetic properties of geometrically frustrated vanadium spinels}
\author{Sohan Lal and Sudhir K. Pandey}
\affiliation{School of Engineering, Indian Institute of Technology Mandi, Kamand 175005, Himachal Pradesh, India}

\date{\today}

\maketitle

\section{Abstruct}

The inconsistency about the degree of geometrical frustration has been a long issue in AV$_{2}$O$_{4}$ (A $\equiv$ Zn, Cd and Mg) compounds, which arises from the two experimental results: (i) frustration indices and (ii) magnetic moments. In the present study, we try to understand such inconsistency by using {\it ab initio} electronic structure calculations. The orbital degrees of freedom are found to play an important role in understanding the geometrically frustrated magnetic behaviour of these compounds. The magnitude of the maximum calculated values of orbital magnetic moment per formula unit for ZnV$_{2}$O$_{4}$, MgV$_{2}$O$_{4}$ and CdV$_{2}$O$_{4}$ compounds are found to be $\sim$1.54 $\mu$$_{B}$, $\sim$0.92 $\mu$$_{B}$ and $\sim$1.74 $\mu$$_{B}$, respectively. The inclusion of the orbital and spin angular momenta for calculating the frustration indices improves the understanding about the degree of geometrical frustration in these compounds. The calculated values of the frustration indices ($f$$_{\it J}$) are largest for MgV$_{2}$O$_{4}$ and smallest for CdV$_{2}$O$_{4}$ for 3.3$\leq$ $U \leq$5.3 eV. In this range of $U$, the calculated values of $\Delta$M$_{2}$=M$_{\rm total}$-M$_{\rm exp}$ (where, M$_{\rm total}$=M$_{\rm spin}$-$\arrowvert$M$_{\rm orbital}$$\arrowvert$) are also found to be largest for MgV$_{2}$O$_{4}$ and smallest for CdV$_{2}$O$_{4}$. Hence, the consistency about the degree of geometrical frustration, which arises from the $f$$_{\it J}$ as well as from the $\Delta$M$_{2}$ is achieved and improves the understanding about the degree of geometrical frustration in these compounds. Calculated values of band gap in this range of $U$ are found to be closer to that of experimentally observed values for all three compounds. The absolute values of the nearest neighbour exchange coupling constant ({\it J$_{nn}$}) between V spins are found to be largest for MgV$_{2}$O$_{4}$ and smallest for CdV$_{2}$O$_{4}$, which indicate that the calculated absolute values of the Curie-Weiss temperature ($\varTheta$$_{CW}$)$_{\it J}$ are highest for MgV$_{2}$O$_{4}$ and smallest for CdV$_{2}$O$_{4}$ for 3.3$\leq$ $U \leq$5.3 eV. In this range of $U$, the magnetic transition temperature ($T$$_{N}$)$_{\it J}$ is found to be $\sim$150 K, $\sim$60 K and $\sim$22 K for MgV$_{2}$O$_{4}$, ZnV$_{2}$O$_{4}$ and CdV$_{2}$O$_{4}$, respectively, which shows that the order of ($T$$_{N}$)$_{\it J}$ is similar to that of ($T$$_{N}$)$_{\rm exp}$ for these compounds. Hence, all the magnetic properties studied in the present work are well explained in these spinels for 3.3$\leq$ $U \leq$5.3 eV. This work is expected to provide a valuable input in understanding the geometrically frustrated magnetic behaviour for those systems for which the orbital part of the angular momenta are not quenched.

\section{Introduction}

    The ground state properties in the condensed matter physics have been well described from long time by the density functional theory (DFT) approach.\cite{Lundqvist,Mahan,Czyzyk} At least in the metallic systems, such properties are described well either by local density approximation (LDA) or generalized gradient approximation (GGA) based on the DFT. However, it is well known that these functionals underestimate the orbital moments for strongly correlated systems, which are induced by the strong spin orbit coupling (SOC).\cite{Kraft,Brooks,Oppeneer,Bultmark} This can be resolved by adding orbital dependent Hartree-Fock (HF) potential to the LDA/GGA, so called LDA+{\it U}/LDA+{\it U} approximation.\cite{Bultmark,Liechtenstein,Solovyev,Anisimov} A major problem in this approach is that the electron-electron interaction has already been included in LDA/GGA potential. Hence, the simple addition of HF potential leads to double counting (DC). Here, the best way is to identify the mean-field-part of the HF potential and subtract it, leaving the orbital dependent correction to the mean field type LDA potential. Czy$\dot{\rm z}$yk and Sawatzky suggested a scheme that is true for uniform occupancies, so called around mean field (AMF) and is applicable to the weakly correlated systems.\cite{Czyzyk} For strongly correlated systems, AMF DC scheme is not valid due to the lack of uniform occupancies. For these systems, one can prefer the fully-localized (FL) DC scheme, where the average effect for a localized state is subtracted with integer occupation number.\cite{Liechtenstein,Bultmark}
     
    Geometrical frustration is always a challenging problem in the strongly correlated systems. In these systems it arises due to the interactions between spin degrees of freedom in a lattice, which are incompatible to that of the essential crystal geometry. In the highly geometrically frustrated magnets, frustration suppresses the long-range magnetic order and leads to a degenerate manifold of ground states. Degeneracy in the frustrated magnetic systems yield different complex ordering structures, spin liquid states and spin ice states.\cite{Diep,Liebmann} Charge ordering phenomena are also much affected by the geometrical frustration.\cite{Anderson} After 1980's, spin systems on the pyrochlore lattice (an example of geometrically frustrated structures) have been studied in more detail.\cite{Liebmann,Canals,Harris,Tsunetsugu2001,Koga,Berg,Reimers,Moessner} In the classical Heisenberg spin systems, a very strong geometrical frustration is anticipated due to the antiferromagnetically coupled spins on a pyrochlore lattice, which do not show long-range order at any nonzero temperature.\cite{Liebmann,Reimers,Moessner}          
                      
    Vanadium spinels, AV$_{2}$O$_{4}$ (A $\equiv$ Zn, Cd and Mg) with the face-centered-cubic structure at room temperature are an interesting example of geometrically frustrated systems, which show a variety of physical properties.\cite{Tsunetsugu,Tchernyshyov,Matteo,Motome,Maitra,Giovannetti,Pandey2011,Pandey2012,Lal,Lee,Suzuki} In these systems a pyrochlore lattice is formed due to the corner sharing network of tetrahedra with magnetically coupled V atoms that gives the geometrical frustration.\cite{Lee,Suzuki,Mamiya,Reehuis,Onoda,Adachi,Garlea,Chung} As opposed to the other geometrically frustrated systems, vanadium spinels show the long range antiferromagnetic ordering at low temperature.\cite{Gardner} In these compounds, orbital ordering is found to be responsible for the structural transition that leads to a long range magnetic ordering due to the lifting of geometrical frustration in the tetragonal structure.\cite{Huang,Radaelli} The structural transition temperature ($T$$_{S}$) for CdV$_{2}$O$_{4}$ ($\sim$97 K)$>$MgV$_{2}$O$_{4}$ ($\sim$65 K)$>$ZnV$_{2}$O$_{4}$ ($\sim$50 K) compound.\cite{Mamiya,Nishiguchi,Reehuis,Onoda,Radaelli,Wheeler,Takagi} However, the magnetic transition temperature ($T$$_{N}$) for MgV$_{2}$O$_{4}$ ($\sim$42 K)$>$ZnV$_{2}$O$_{4}$ ($\sim$40 K)$>$CdV$_{2}$O$_{4}$ ($\sim$35 K) compound.\cite{Mamiya,Reehuis,Onoda,Nishiguchi,Radaelli,Wheeler,Takagi} The values of $T$$_{N}$ for these compounds are always found to be less than $T$$_{S}$. However, it is expected that the values of $T$$_{S}$ and $T$$_{N}$ should be same for these compounds because the geometrical frustration is removed by the above proposed mechanism. The fact that $T$$_{N}$$<$$T$$_{S}$ in these compounds, indicates the presence of certain degree of geometrical frustration in the tetragonal phase of the compound.  
     
    The presence of geometrical frustration in the above mentioned compounds can be attributed to the following two experimental facts. First one is the frustration index, defined as $f$$_{\it S}$=$\arrowvert$$\varTheta$$_{CW}$$\arrowvert$/$T$$_{N}$, where $\arrowvert$$\varTheta$$_{CW}$$\arrowvert$ and $T$$_{N}$ are the Curie-Weiss temperature and the magnetic ordering temperature, respectively. Experimentally reported values of $\varTheta$$_{CW}$ for ZnV$_{2}$O$_{4}$, MgV$_{2}$O$_{4}$ and CdV$_{2}$O$_{4}$ are $\sim$850 K $\sim$600 K and $\sim$400 K, respectively.\cite{Takagi,Niziol} However, in some literature these values are reported to be different for these compounds depending upon the experimental conditions.\cite{Reehuis,Ueda,Mamiya1,Zhang,Baltzer} In general, one can use the parameter $f$$_{\it S}$ as a measure of strength of geometrical frustration in the system.\cite{Takagi,Pandey2011,Gardner} The values of $f$$_{\it S}$ for ZnV$_{2}$O$_{4}$, CdV$_{2}$O$_{4}$ and MgV$_{2}$O$_{4}$ are found to be 21.3, 11.4 and 14.3, respectively.\cite{Takagi} Among these compounds, ZnV$_{2}$O$_{4}$ is largest frustrated and CdV$_{2}$O$_{4}$ is least frustrated as large value of $f$$_{\it S}$ indicates the more frustration in the system. The second one is based on the difference between the spin magnetic moment per formula unit (MM/f.u.) [M$_{\rm spin}$] and the experimentally observed MM/f.u. (M$_{\rm exp}$), which is denoted by $\Delta$M$_{1}$=M$_{\rm spin}$-M$_{\rm exp}$ in the antiferromagnetic phase of these compounds. M$_{\rm spin}$ for these compounds is found to be 4 $\mu$$_{B}$. However, M$_{\rm exp}$ for CdV$_{2}$O$_{4}$, ZnV$_{2}$O$_{4}$ and MgV$_{2}$O$_{4}$ compounds are reported to be 2.38 $\mu$$_{B}$, 1.26 $\mu$$_{B}$ and 0.94 $\mu$$_{B}$, respectively.\cite{Lee,Reehuis,Onoda,Wheeler} Large values of $\Delta$M$_{1}$ is explained in two different ways. One way of understanding is to attribute this difference to the activeness of geometrical frustration in these compounds, which results in the quantum fluctuations responsible for the reduction of M$_{\rm exp}$ drastically as reported in MgV$_{2}$O$_{4}$.\cite{Wheeler} If this is the case, then the degree of the geometrical frustration for MgV$_{2}$O$_{4}$$>$ZnV$_{2}$O$_{4}$$>$CdV$_{2}$O$_{4}$. Also, the order of geometrical frustration that arises from the  experimentally reported frustration index is found to be ZnV$_{2}$O$_{4}$$>$MgV$_{2}$O$_{4}$$>$CdV$_{2}$O$_{4}$ as described in the reference [43]. Hence, both the results show the inconsistency about the degree of geometrical frustration in these compounds. Another way of understanding will be by considering contribution of the magnitude of orbital part of MM/f.u. (denoted by $\arrowvert$M$_{\rm orbital}$$\arrowvert$) reported by Maitra {\it et al.} for ZnV$_{2}$O$_{4}$.\cite{Maitra} The calculated values of the $\arrowvert$M$_{\rm orbital}$$\arrowvert$ for ZnV$_{2}$O$_{4}$, MgV$_{2}$O$_{4}$ and CdV$_{2}$O$_{4}$ compounds are found to be $\sim$1.50 $\mu$$_{B}$, $\sim$0.4 $\mu$$_{B}$ and $\sim$0.4 $\mu$$_{B}$, respectively.\cite{Maitra,Pandey2011,Pandey2012} Even by considering the $\arrowvert$M$_{\rm orbital}$$\arrowvert$, the total calculated MM/f.u. (M$_{\rm total}$) is still larger than the M$_{\rm exp}$, indicating that the certain degree of geometrical frustration is still present in these compounds. The inclusion of $\arrowvert$M$_{\rm orbital}$$\arrowvert$ does not improve the situation. 
      
    The above discussion clearly suggests that the current understanding of these compounds is not sufficient to resolve the contradiction about the degree of geometrical frustration present especially in ZnV$_{2}$O$_{4}$ and MgV$_{2}$O$_{4}$. In the present work we have tried to resolve this contradiction by carrying out detailed LDA+$U$+SOC calculations. Our results clearly show the importance of orbital degrees of freedom along with the spin in understanding the magnetic properties of these compounds. The inclusion of orbital and spin angular momenta for calculating the frustration indices improves the understanding about degree of geometrical frustration present in these compounds. The calculated values of the frustration indices $f$$_{\it J}$ thus obtained for 3.3$\leq$ $U \leq$5.3 eV provide the largest value for MgV$_{2}$O$_{4}$ and smallest for CdV$_{2}$O$_{4}$. This behaviour is similar to that expected from the $\Delta$M$_{2}$. In this range of $U$, the magnitude of Curie-Weiss temperature [($\varTheta$$_{CW}$)$_{\it J}$] are highest for MgV$_{2}$O$_{4}$ and smallest for CdV$_{2}$O$_{4}$, which indicate that the magnitude of the nearest neighbour exchange coupling constant ({\it J$_{nn}$}) among V spins are largest for MgV$_{2}$O$_{4}$ and smallest for CdV$_{2}$O$_{4}$ compounds. Also, the order of magnetic transition temperature [($T$$_{N}$)$_{\it J}$] is found to be consistent with the experimentally reported order of magnetic transition temperature ($T$$_{N}$)$_{\rm exp}$ for these spinels for above mentioned range of $U$.
   
\section{Computational details}

    In present work, the ferromagnetic (FM) and antiferromagnetic (AFM) electronic-structure calculations of AV$_{2}$O$_{4}$ (A $\equiv$ Zn, Cd and Mg) compounds are carried out by using the {\it state-of-the-art} full-potential linearized augmented plane wave (FP-LAPW) method.\cite{elk} The atomic positions and lattice parameters used in the calculations for every compounds are taken from the literature.\cite{Onoda,Reehuis,Wheeler} All these calculations are performed in the tetragonal phase for which Perdew -Wang/Ceperley -Alder exchange correlation functional has been used.\cite{Perdew} In order to calculate the nearest neighbour exchange coupling constant ({\it J$_{nn}$}), we have considered the two magnetic orderings: FM and AFM (not the experimentally observed structure) ordering of the spins of the four V atoms in a primitive unit cell. For AFM ordering, the primitive unit cell consists of two up and two down spins on the four V atoms. The effect of on-site Coulomb interaction among V 3$d$ electrons is considered within LDA+$U$ formulation of the density functional theory.\cite{Bultmark} Normally in this method $U$ and $J$ are used as parameters. However in our calculations, only $U$ is used as a free parameter and the value of $J$ is calculated self-consistently as described in the reference [7]. FL and AMF DC schemes have been used in these calculations.\cite{Liechtenstein,Czyzyk} Both the DC schemes give the similar results. Thus in order to avoid the repetition of the results, we have discussed the results corresponding to the FL DC scheme in the rest of the manuscript. The  muffin-tin sphere radii used in the present work are 2.0, 2.46, 1.39, 2.0 and 1.54 Bohr for Zn, Cd, Mg, V and O, respectively. (6,6,6) k-point mesh size is used in the calculations. In order study the role of orbital degrees of freedom, SOC is included self-consistently in LSDA+$U$ calculations, where no external parameters have been used. The effect of the spin canting has been found to be small for these spinels both theoretically and experimentally.\cite{Wheeler, Pandey2011, Pandey2012} Hence, collinear calculations have been performed here by varying $U$ from 2-6 eV, where the direction of the magnetization is set along the z-axis. In this range of $U$, every calculations are started from the different combinations for converged electron densities and potentials corresponding to different values of $U$ for every compounds. The convergence target of total energy has been achieved below ~10$^{-4}$ Hartrees/cell.

\section{Results and discussions}

    The crystal structure and splitting of $d$ levels of V ion of AV$_{2}$O$_{4}$ (A $\equiv$ Zn, Cd and Mg) compounds in presence of octahedral crystal field and SOC are shown in the Fig. 1(a and b). It is clear from the Fig. 1(a) that each V ion is located at the center of the edge sharing octahedra which create a crystal field. Such a crystal field splits the $d$ levels into lower energy t$_{2g}$ levels and higher energy e$_{g}$ levels in the cubic phase as shown in the Fig. 1(b). Because of ignoring the small trigonal distortion present in these systems, these t$_{2g}$ levels are  normally assumed to be threefold degenerate. The t$_{2g}$ levels are generally represented by orbital angular momentum $l$=1. The degeneracy of t$_{2g}$ levels is further lifted by the SOC, where the lower value of total angular momentum $j$=1/2 corresponds to the lower energy state.             
    
     In order to see the effect of SOC for all the three compounds, we have first calculated the SOC energy (the energy difference between LSDA+$U$+SOC and LSDA+$U$ energies for these compounds) roughly. Here, as a representative, we have shown the calculated value of SOC energy only for $U$=4 eV. The magnitude of rough estimated SOC energy per formula unit for ZnV$_{2}$O$_{4}$, CdV$_{2}$O$_{4}$ and MgV$_{2}$O$_{4}$ are $\sim$0.5 eV, $\sim$0.5 eV and $\sim$0.2 eV, respectively for $U$=4 eV. Now, in order to compare the calculated MM to the experimentally observed MM, we have used the FM structure instead of experimentally observed AFM structure in the present calculations. This is because of the following reasons. It is important to note that the various physical quantities ($f$$_{\it S}$, $\varTheta$$_{CW}$ and $T$$_{N}$) for these compounds are observed experimentally for spin, $S$=1. Now in order to compare our results with the experimental data (discussed later of the manuscript), FM structure calculations are best for calculating the MM as compared to the experimentally observed AFM structure calculations for these compounds. This is because of the fact that in AFM calculations, the total MM in the interstitial region is almost zero and hence the M$_{\rm spin}$ (comes only inside the muffin-tin spheres) is always less than 4.0 $\mu$$_{B}$ for $U$=2-6 eV, which indicates that $S$ $<$1. However, in FM calculations values of M$_{\rm spin}$ comes out to be 4.0 $\mu$$_{B}$ (both inside the muffin-tin spheres as well as from the interstitial region) as per expectation for whole range of $U$, which indicates that the V atom is in 3+ ionic state with spin, $S$=1. However, the orbital MM is calculated only inside the muffin-tin spheres because there is no way of calculating the orbital MM in the interstitial region. It is important to note that with increase in the muffin-tin sphere radius of V atom more than that used in the present calculations, only a slight changes in the orbital MM has been observed, which indicates that the contribution of orbital part of MM is almost negligible from the interstitial region. Also for calculating M$_{\rm total}$ (comes from both spin and orbital part of MM), we have also calculated the $\arrowvert$M$_{\rm orbital}$$\arrowvert$. 
     
    The plot of MM/f.u. as a function of $U$ is shown in the Fig. 2(a-c). It is evident from the figure that the values of the M$_{\rm orbital}$ are negative for these compounds, which indicate that the direction of the orbital MM is opposite to the spin MM consistent with the Hund's rule for less than half-filled cases. The behavior of MM/f.u. for ZnV$_{2}$O$_{4}$ is similar (different) to that of MgV$_{2}$O$_{4}$ (CdV$_{2}$O$_{4}$) for whole range of $U$ studied here. The $\arrowvert$M$_{\rm orbital}$$\arrowvert$ shows a small dependence on $U$ from 2 to 3 eV for both ZnV$_{2}$O$_{4}$ and MgV$_{2}$O$_{4}$ compounds. In this range of $U$ both compounds gives almost equal values of $\arrowvert$M$_{\rm orbital}$$\arrowvert$, where it increases from $\sim$0.10 to $\sim$0.20 $\mu$$_{B}$. However, it increases sharply to $\sim$1.52 (0.78) $\mu$$_{B}$ for ZnV$_{2}$O$_{4}$ (MgV$_{2}$O$_{4}$) compound at $U$=3.5 eV. The value of $\arrowvert$M$_{\rm orbital}$$\arrowvert$ remains almost same for ZnV$_{2}$O$_{4}$ and increases to $\sim$0.92 $\mu$$_{B}$ for MgV$_{2}$O$_{4}$ as $U$ changes from 3.5 to 4 eV. For CdV$_{2}$O$_{4}$, it increases from $\sim$0.82 to $\sim$1.74 $\mu$$_{B}$ as $U$ varies from 2 to 3.5 eV.  For $U \geq$4 eV, it decreases continuously upto $\sim$0.68 $\mu$$_{B}$, $\sim$0.42 $\mu$$_{B}$ and $\sim$0.62 $\mu$$_{B}$ for ZnV$_{2}$O$_{4}$, MgV$_{2}$O$_{4}$ and CdV$_{2}$O$_{4}$, respectively at $U$=6 eV. The possible cause of changes of $\arrowvert$M$_{\rm orbital}$$\arrowvert$ by varying $U$ observed in these compounds is discussed below as: In regular octahedron symmetry five fold degenerate $d$ orbitals split into lower energy triply degenerate t$_{2g}$ levels and higher energy doubly degenerate e$_{g}$ levels. For large crystal field splitting, the lower energy t$_{2g}$ levels are normally represented by angular momentum $l$=1 ($l^{z}$=-1, 0, +1) and higher energy e$_{g}$ levels by $l$=$\frac{1}{2}$ ($l^{z}$=-$\frac{1}{2}$, +$\frac{1}{2}$). In tetragonal distortion with $c/a <$1, t$_{2g}$ levels are split into lower energy $d_{xy}$ ($l^{z}$=0) orbital and higher energy degenerate $d_{xz}$ and $d_{yz}$ ($l^{z}$=-1, +1) orbitals. Similarly, e$_{g}$ levels split into $d_{x^{2}-y^{2}}$ and $d_{z^{2}}$ orbitals.\cite{Khomskii,Kugel,Wheeler} Thus the total orbital angular momentum ($L$) of V ions are decided by the number of electrons in each of these five $d$ orbitals. For example, if we consider pure ionic picture along with full orbital polarization for these compounds, then one out of two V 3$d$ electrons occupies the $d_{xy}$ orbital and second electron occupies either $d_{xz}$ or $d_{yz}$ orbital depending upon the site as these compounds show orbital ordering. Hence, result $L$=1 for $d$ electrons. However, presence of certain degree of hybridization breaks the fully ionic picture of the compounds and 3$d$ electrons may occupy all the five $d$ orbitals depending upon crystal field splitting energy, Hunds coupling energy, on-site Coulomb interaction, etc. These electrons occupancies of various orbitals decide the $L$ of the V ions. This discussion clearly suggests that the localization of electrons (which changes the total number of $d$ electrons at V sites) along with their distribution in various $d$ orbitals finally decide the values of $L$ for V ion. Hence, with increasing $U$ from 2 to 4 eV, the total number of $d$ electrons (means fractional occupation of all five $d$ orbitals) of V ion increases, which result the increase in $\arrowvert$M$_{\rm orbital}$$\arrowvert$ because of the more localization of the electrons as per expectation.\cite{Liechtenstein,Anisimov,Solovyev} Above U=4 eV, one can expect that the decrease in $\arrowvert$M$_{\rm orbital}$$\arrowvert$ is due to the delocalization of electrons. But this is not the case. This is because for $U \geq$4 eV, the total number of $d$ electrons of V ion remains constant upto $U$=6 eV. However, the redistribution of electrons in all five $d$ orbitals above $U$=4 eV have been observed in the density matrix, which is expected to be responsible for decrease in $\arrowvert$M$_{\rm orbital}$$\arrowvert$. This type of non-monotonic behavior of $\arrowvert$M$_{\rm orbital}$$\arrowvert$ with increasing $U$ from 2 to 6 eV is expected in these compounds as shown by Bultmark $et$ $al.$ in ferromagnetic PuP and US systems.\cite{Bultmark} The maximum calculated value of $\arrowvert$M$_{\rm orbital}$$\arrowvert$ for ZnV$_{2}$O$_{4}$ is found to be $\sim$1.54 $\mu$$_{B}$  at $U$=4 eV, which is in good agreement with that calculated by Maitra {\it et al.} as shown in the Table 1.\cite{Maitra} However, it is found to be $\sim$0.92 $\mu$$_{B}$ at $U$=4 eV and $\sim$1.74 $\mu$$_{B}$ at $U$=3.5 eV for MgV$_{2}$O$_{4}$ and CdV$_{2}$O$_{4}$, respectively and is larger than that predicted by different groups as also shown in the Table 1.\cite{Pandey2011,Pandey2012,Kaur} The value of $\arrowvert$M$_{\rm orbital}$$\arrowvert$ depends on the exchange correctional functional, on-site exchange parameter $J$, muffin-tin sphere radii and type of calculations (collinear or non-collinear), which are different in the present calculations as compared to the earlier calculations. From above discussion, it is clear that the calculated values of $\arrowvert$M$_{\rm orbital}$$\arrowvert$ are different for these compounds for whole range of $U$ studied here. This is due to the fact that the distorted VO$_{6}$ octahedron contains V-O bonds and V-O-V angles, which are different for these compounds and are partly attributed to the difference in the ionic radii at the A=Cd$^{2+}$, Mg$^{2+}$ and Zn$^{2+}$ site. Therefore, the values of $\arrowvert$M$_{\rm orbital}$$\arrowvert$ are expected to be different for all three compounds because it comes from the distribution of electrons among the V 3$d$ orbitals. Such a distribution of electrons in these orbitals are greatly influenced by the V-O bonds and V-O-V angles. It is evident from the figure that the contribution of the $\arrowvert$M$_{\rm orbital}$$\arrowvert$ to the M$_{\rm total}$ (where, M$_{\rm total}$=M$_{\rm spin}$-$\arrowvert$M$_{\rm orbital}$$\arrowvert$) is more for CdV$_{2}$O$_{4}$ as compared to ZnV$_{2}$O$_{4}$ and MgV$_{2}$O$_{4}$ compounds for $U$=2-4 eV. Above $U$=4 eV, its contribution to the M$_{\rm total}$ is largest for ZnV$_{2}$O$_{4}$ and smallest for MgV$_{2}$O$_{4}$ compound. In these compounds, the values of M$_{\rm total}$ show the similar behavior to that of M$_{\rm orbital}$ for whole range of $U$. For $U \leq$3 eV, it is almost equal to the M$_{\rm spin}$ for ZnV$_{2}$O$_{4}$ and MgV$_{2}$O$_{4}$ compounds, which is due to the small contribution of $\arrowvert$M$_{\rm orbital}$$\arrowvert$. However, for CdV$_{2}$O$_{4}$, it decreases from $\sim$3.18 to $\sim$2.26 $\mu$$_{B}$ as $U$ changes from 2-3.5 eV. For ZnV$_{2}$O$_{4}$ (MgV$_{2}$O$_{4}$) compound, it decreases sharply to $\sim$2.48 ($\sim$3.22) $\mu$$_{B}$ at $U$=3.5 eV. The value of $\arrowvert$M$_{\rm total}$$\arrowvert$ remains almost same for ZnV$_{2}$O$_{4}$ and decreases to $\sim$3.08 $\mu$$_{B}$ for MgV$_{2}$O$_{4}$ as $U$ changes from 3.5 to 4 eV. For $U \geq$4 eV, it increases continuously upto $\sim$3.32 $\mu$$_{B}$, $\sim$3.58 $\mu$$_{B}$ and $\sim$3.38 $\mu$$_{B}$ for ZnV$_{2}$O$_{4}$, MgV$_{2}$O$_{4}$ and CdV$_{2}$O$_{4}$, respectively at $U$=6 eV. Here, it is interesting to compare the calculated values of M$_{\rm total}$ to that of experimentally observed values of M$_{\rm exp}$ for these compounds, which are also shown in the Table 1. The value of M$_{\rm exp}$ has been taken from the references [30,33,34,41]. The calculated values of M$_{\rm total}$ is larger than M$_{\rm exp}$ for ZnV$_{2}$O$_{4}$ (1.26 $\mu$$_{B}$) and MgV$_{2}$O$_{4}$ (0.94 $\mu$$_{B}$) compounds for all values of $U$. However, for CdV$_{2}$O$_{4}$ compound, the calculated values of M$_{\rm total}$ is larger than the M$_{\rm exp}$ (2.38 $\mu$$_{B}$) for $U$ $<$3 eV and $U$ $>$4 eV. For 3$\leq U$ $\leq$4 eV, the values of M$_{\rm total}$ is almost equal to M$_{\rm exp}$. The value of M$_{\rm exp}$$<$M$_{\rm total}$ suggests the presence of certain degree of geometrical frustration in these compounds. 

    In order to make this point more clear, we have plotted the difference between the M$_{\rm total}$ and M$_{\rm exp}$ ($\Delta$M$_{2}$=M$_{\rm total}$-M$_{\rm exp}$) in the Fig. 3. It is clear from the figure that the behavior of $\Delta$M$_{2}$ for both ZnV$_{2}$O$_{4}$ and MgV$_{2}$O$_{4}$ compounds are almost similar for whole range of $U$. For ZnV$_{2}$O$_{4}$ (MgV$_{2}$O$_{4}$), $\Delta$M$_{2}$ decreases slowly from $\sim$2.64 ($\sim$2.96) to $\sim$2.54 ($\sim$2.88) $\mu$$_{B}$ when $U$ changes from 2 to 3 eV. At $U$=3.5 eV, it decreases sharply to $\sim$1.22 ($\sim$2.28) $\mu$$_{B}$ for ZnV$_{2}$O$_{4}$ (MgV$_{2}$O$_{4}$) and remains almost constant upto $U$=4 eV. Above $U$=4 eV, it increases continuously upto $\sim$2.06 ($\sim$2.64) $\mu$$_{B}$ for ZnV$_{2}$O$_{4}$ (MgV$_{2}$O$_{4}$) at $U$=6 eV. For CdV$_{2}$O$_{4}$, $\Delta$M$_{2}$ decreases continuously from $\sim$0.80 to $\sim$0.0 $\mu$$_{B}$ as $U$ varies from 2 to 4 eV and then increases upto $\sim$1.0 $\mu$$_{B}$ at $U$=6 eV. It is also clear from the figure that the $\Delta$M$_{2}$ for MgV$_{2}$O$_{4}$$>$ZnV$_{2}$O$_{4}$$>$CdV$_{2}$O$_{4}$ for all values of $U$. Here, we attribute $\Delta$M$_{2}$ to the degree of activeness of geometrical frustration in these compounds which results in the quantum fluctuations responsible for the reduction of MM. Hence among these compounds, the degree of geometrical frustration is largest for MgV$_{2}$O$_{4}$ and smallest for CdV$_{2}$O$_{4}$. From above discussion, it is clear that the order of $\Delta$M$_{2}$ is similar as that of $\Delta$M$_{1}$ for these compounds, which indicates that even by including the M$_{\rm orbital}$ to the M$_{\rm total}$ does not change the order of degree of geometrical frustration. Hence, the degree of geometrical frustration, which arises from $\Delta$M$_{2}$ is inconsistent to that observed from the experimentally reported $f$$_{\it S}$. 
         
     It is important to note that the experimentally reported values of $f$$_{\it S}$ in these compounds are calculated by the ratio between the Curie-Weiss temperature and the antiferromagnetic transition temperature as,

\begin{equation}     
f_{\it S}=\arrowvert(\varTheta_{CW})_{\rm exp}\arrowvert/(T_{N})_{\rm exp}
\end{equation}   

The values of ($\varTheta$$_{CW}$)$_{\rm exp}$ are calculated (without including the orbital angular momentum $L$) by the following formula,\cite{Ashcroft}

\begin{equation}     
(\varTheta_{CW})_{\it exp}=\frac{N(g_{S}\mu_{B})^{2}}{3k_{B}V}S(S+1)\lambda
\end{equation}     

Where $N/V$, $g$$_{S}$, $\mu$$_{B}$, $k$$_{B}$, $S$ and $\lambda$ are the number of magnetic atoms per unit volume, Lande g factor for the total spin angular momentum, Bohr magneton, Boltzmann constant, total spin angular momentum and Weiss molecular field constant, respectively. Substituting Eqn. (2) in Eqn. (1), we get
\begin{equation}
       f_{\it S}=\frac{8N\mu_{B}^{2}}{3k_{B}V(T_{N})_{\rm exp}}\lambda 
\end{equation}
for $S$=1 and $g$$_{S}$=2.

For all three compounds, we have seen above that the contribution from the orbital part of MM is significant and can not be neglected. So, the above approximation is not expected to give correct values of the frustration indices. The best way for calculating the frustration indices is to include the contribution from both orbital and spin angular momenta. Now in order to calculate the frustration indices for these compounds, we have replaced the total spin angular momentum $S$ by the total angular momentum $J$ in Eqn. (2) and after using Eqn. (1), we get
    
\begin{equation}
       f_{\it J}=\frac{N(g_{J}\mu_{B})^{2}}{3k_{B}V(T_{N})_{\rm exp}}J(J+1)\lambda 
\end{equation}
    
Where $f$$_{\it J}$, $J$=$\arrowvert$$L$-$S$$\arrowvert$, $L$ and $g$$_{J}$=1+$\frac{J(J+1)+S(S+1)-L(L+1)}{2J(J+1)}$ are the frustration index (by including both orbital and spin angular momenta), total angular momentum for less than half filled $d$ orbitals, total orbital angular momentum and Lande g factor for the total angular momentum, respectively. Dividing Eqn. (4) by Eqn. (3), we get

\begin{equation}
       f_{\it J}=\frac{g^{2}_{J}J(J+1)}{8}f_{\it S} 
\end{equation}
The experimentally observed values of $f$$_{\it S}$ are given in the Table 1.

    Now, using Eqn. (5), we have calculated the values of $f$$_{\it J}$ by including both orbital and spin angular momenta for these compounds by varying $U$ from 2 to 6 eV. In Fig. 4, we have plotted the $f$$_{\it J}$ for these compounds as a function of $U$. It is evident from the figure that the values of $f$$_{\it J}$ are largest for ZnV$_{2}$O$_{4}$ and smallest for CdV$_{2}$O$_{4}$, for $U$ $<$3.3 eV. However, for 3.3 $\leq$ $U \leq$ 5.3 eV, the order of $f$$_{\it J}$ is MgV$_{2}$O$_{4}$$>$ZnV$_{2}$O$_{4}$$>$CdV$_{2}$O$_{4}$. For $U$ $>$5.3 eV, the order of $f$$_{\it J}$ in these compounds is similar to that observed for $U$ $<$3.3 eV. It is also clear from the figure that the values of $f$$_{\it J}$ for CdV$_{2}$O$_{4}$ are always found to be less than ZnV$_{2}$O$_{4}$ and MgV$_{2}$O$_{4}$ compounds for whole range of $U$, which indicates that it is least frustrated among these spinels. The reason for the different values of $f$$_{\it J}$ for these compounds for whole range of $U$ is discussed here as. It is also clear from the Eqn. (5) that the values of $f$$_{\it J}$ depends on both the contribution of $L$ to the $J$ as well as the values of $f$$_{\it S}$. The reason for the different values of $L$ for these compounds as a function of $U$ has been already discussed above. Here, we discuss only for $f$$_{\it S}$. In the tetragonal phase, distorted VO$_{6}$ octahedron and V$_{4}$ tetrahedron have different V-O-V angles and V-V bonds, respectively, which are different for these compounds. Because of the different values of these bonds and angles from compound to compound (means CdV$_{2}$O$_{4}$ to MgV$_{2}$O$_{4}$ to ZnV$_{2}$O$_{4}$), the values of $f$$_{\it S}$ is expected to be different for whole range of $U$ studied here. For, example the strength of the magnetic interactions (or $\arrowvert$$\varTheta$$_{CW}$$\arrowvert$) are small for CdV$_{2}$O$_{4}$ as compared to MgV$_{2}$O$_{4}$ and ZnV$_{2}$O$_{4}$ because of the large V-V bonds and V-O-V angles in CdV$_{2}$O$_{4}$ as compared to other two compounds. Also, Canosa $et$ $al.$ have proposed that the $T$$_{N}$ of these compounds also increases with decreasing V-V distances.\cite{Canosa} Hence, geometrical frustration is expected to different for these compounds because of the different values of $\arrowvert$$\varTheta$$_{CW}$$\arrowvert$, $T$$_{N}$ and $L$. Here, it is interesting to compare the order of calculated values of $f$$_{\it J}$ with the calculated values of $\Delta$M$_{2}$ as both are attributed as the degree of geometrical frustration in these compounds. The values of $f$$_{\it J}$ calculated for $U$=3.3-5.3 eV provide the correct ordering of frustration indices, which is consistent with that arises from the $\Delta$M$_{2}$. Hence, in this range of $U$, a long issue of the inconsistency about the degree of geometrical frustration has resolved, which indicate the importance of orbital degrees of freedom in these compounds. This range of $U$ also appear to make sense in the light of the work of Canosa \textit{et al.}, where $d$ electrons of CdV$_2$O$_4$ is put in the localized-electron regime; and ZnV$_2$O$_4$ and MgV$_2$O$_4$ are put in intermediate localization ragime.\cite{Canosa} At that point, it is important to note that the calculated values of band gap in this range of $U$ are found to be closer to that of experimentally observed values for all three compounds.   

    Now, in order to evaluate the nearest neighbour exchange coupling constant {\it J$_{nn}$} between the V atoms of above mentioned compounds, we have considered the FM and AFM (not the experimentally observed structure) spin configurations. This is because of the fact that the experimentally observed sign of exchange coupling constant {\it J$_{nn}$} gives the information about the magnetic ground state of the compound and does not explain it's spin structure. Lowest energy spin structure (i.e. magnetic ground state) is obtained by comparing various spin configurations of the compound. Hence, the FM and AFM primitive unit cell contains four V atoms (V1V2V3V4) with spin $\uparrow$$\uparrow$$\uparrow$$\uparrow$ and $\downarrow$$\downarrow$$\uparrow$$\uparrow$ (used in the present calculations), respectively are sufficient to estimate the {\it J$_{nn}$} for all three compounds. The atomic and spin arrangements of all four V atoms in the AFM primitive unit cell of these spinels are shown in the Fig. 5. Here, we have calculated the total energies of FM and AFM ordering in the classical Heisenberg Hamiltonian of the following form:\cite{Ashcroft}
    
    \begin{equation}
       {\it H} = -{\it J_{nn}}\sum_{\it i>j} S_{i}.S_{j} 
    \end{equation} 
             
where, {\it S$_{i}$} and {\it S$_{j}$} are the spin operators of V sites $i$ and $j$, respectively with S = 1 for V 3$d$$^{2}$ electrons. In these calculations, we have considered only nearest neighbour interactions. In spinels, the magnetic unit cell contains six nearest neighbour interactions among V atoms. Hence, the energy of the FM and AFM ordering per unit cell is now expressed as:

     \begin{equation}
       {\it E_{FM}} = {\it E_{0}}-6{\it J_{nn}} 
    \end{equation} and
    
    \begin{equation}
       {\it E_{AFM}} = {\it E_{0}}+2{\it J_{nn}} 
    \end{equation}
     
 After solving these two Eqns., we get
 
  \begin{equation}
     {\it J_{nn}} =  \frac{{\it E_{AFM}}-{\it E_{FM}}}{8} 
    \end{equation}
    
Using Eqn. (9), we have calculated {\it J$_{nn}$} for these compounds by varying U from 2 to 6 eV. The plot of {\it J$_{nn}$} versus $U$ for these spinels is shown in the Fig. 6. It is evident from the figure that the negative values of exchange coupling constant for these compounds indicate the antiferromagnetic interaction for whole range of $U$. For ZnV$_{2}$O$_{4}$ and MgV$_{2}$O$_{4}$ the magnitude of {\it J$_{nn}$} (denoted by $\arrowvert${\it J$_{nn}$}$\arrowvert$) decreases continuously for whole range of $U$ as per expectation if {\it J$_{nn}$}$\approx$$\frac{4t^{2}}{U}$.\cite{Saul} The values of $\arrowvert${\it J$_{nn}$}$\arrowvert$ decreases from $\sim$82.4 ($\sim$90.6) to $\sim$48.7 ($\sim$30.7) meV for MgV$_{2}$O$_{4}$ (ZnV$_{2}$O$_{4}$) as $U$ changes from 2-6 eV. For CdV$_{2}$O$_{4}$, $\arrowvert${\it J$_{nn}$}$\arrowvert$ decreases from $\sim$33.4 to $\sim$5.6 meV as $U$ changes from 2-6 eV. It is also clear from the figure that among these vanadates, the strength of the exchange interaction is largest for the MgV$_{2}$O$_{4}$ and smallest for the CdV$_{2}$O$_{4}$ for $U >$2.5 eV. Now, we have estimated the Curie-Weiss temperature ($\varTheta$$_{CW}$)$_{\it S}$ for these vanadates from the calculated values of {\it J$_{nn}$} by using following formula,\cite{Ashcroft} 

\begin{equation}
   (\varTheta_{CW})_{\it S}=\frac{S(S+1)}{3k_{B}}z{\it J_{nn}}  
    \end{equation}
    
where, $S$=1 (for V 3$d^{2}$ electrons) and z=6 are the total spin angular momentum and the nearest neighbours of V atom (among which the exchange interaction is effective), respectively. Calculated values of ($\varTheta$$_{CW}$)$_{\it S}$ as a function of $U$ for these compounds are given in the Fig. 7. It is clear from the figure that the absolute values of ($\varTheta$$_{CW}$)$_{\it S}$ [denoted by $\arrowvert$($\varTheta$$_{CW}$)$_{\it S}$$\arrowvert$] decreases continuously with increases $U$ from 2 to 6 eV for all compounds. For $U \leq$2.5 eV, it is largest for ZnV$_{2}$O$_{4}$ and smallest for CdV$_{2}$O$_{4}$ compound. As $U$ increases from 2.5 to 6 eV, the order of $\arrowvert$($\varTheta$$_{CW}$)$_{\it S}$$\arrowvert$ becomes MgV$_{2}$O$_{4}$$>$ZnV$_{2}$O$_{4}$$>$CdV$_{2}$O$_{4}$, which indicates that the strength of the magnetic interaction is highest for MgV$_{2}$O$_{4}$ and lowest for CdV$_{2}$O$_{4}$. Here, it is interesting to compare the $\arrowvert$($\varTheta$$_{CW}$)$_{\it S}$$\arrowvert$ with the experimentally observed ($\varTheta$$_{CW}$)$_{\rm exp}$ for 3.3$\leq$ $U \leq$5.3 eV. This is because the consistency about the degree of geometrical frustration (which arises from $f$$_{\it J}$ and $\Delta$M$_{2}$) as well as the the band gap are achieved in this range of $U$ as discussed above. Experimentally reported absolute value of ($\varTheta$$_{CW}$)$_{\rm exp}$ for ZnV$_{2}$O$_{4}$, MgV$_{2}$O$_{4}$ and CdV$_{2}$O$_{4}$ are $\sim$850 K $\sim$600 K and $\sim$400 K, respectively as shown in the Table 1.\cite{Takagi} It is evident from the figure that the $\arrowvert$($\varTheta$$_{CW}$)$_{\it S}$$\arrowvert$ is about 2.5 (4.5) times larger than the experimental one for ZnV$_{2}$O$_{4}$ (MgV$_{2}$O$_{4}$) compounds for 3.3$\leq$ $U \leq$5.3 eV. For CdV$_{2}$O$_{4}$, it is almost close to the ($\varTheta$$_{CW}$)$_{\rm exp}$ for this range of $U$. Here, it is also interesting to see the effect of orbital degrees of freedom on the ($\varTheta$$_{CW}$)$_{\it S}$ because the contribution from the orbital part of MM is significant and can not be neglected for these spinels. Hence, we have replaced the total spin angular momentum $S$ by the total angular momentum $J$ in Eqn. (10), we get 
 \begin{equation}
   (\varTheta_{CW})_{\it J}=\frac{J(J+1)}{3k_{B}}z{\it J_{nn}}  
    \end{equation}
Dividing Eqn. (11) by Eqn. (10), we get

 \begin{equation}
   (\varTheta_{CW})_{\it J}=\frac{J(J+1)}{2} (\varTheta_{CW})_{\it S}
    \end{equation}   

for $S$=1.     
 
Now, Eqn. (12) is used to calculate the ($\varTheta$$_{CW}$)$_{\it J}$, where we have included both orbital and spin angular momenta by varying $U$ from 2 to 6 eV. In Fig. 8, we have plotted the ($\varTheta$$_{CW}$)$_{\it J}$ for these compounds as a function of $U$. From figure, it is clear that the order of absolute values of ($\varTheta$$_{CW}$)$_{\it J}$ [denoted by $\arrowvert$($\varTheta$$_{CW}$)$_{\it J}$$\arrowvert$] is similar to that of $\arrowvert$($\varTheta$$_{CW}$)$_{\it S}$$\arrowvert$ in these compounds for whole range of $U$. For these compounds, the values of $\arrowvert$($\varTheta$$_{CW}$)$_{\it J}$$\arrowvert$ decreases continuously as $U$ changes from 2-4 eV. For $U \geq$ 4 eV, the increase in the $\arrowvert$($\varTheta$$_{CW}$)$_{\it J}$$\arrowvert$ is mainly due to the decrease in the absolute values of orbital angular momenta (as discussed above) for these spinels. The values of ($\varTheta$$_{CW}$)$_{\it J}$ come closer and closer to the ($\varTheta$$_{CW}$)$_{\rm exp}$ for both ZnV$_{2}$O$_{4}$ and MgV$_{2}$O$_{4}$ compounds as $U$ varies from 3.3 to 5.3 eV. For ZnV$_{2}$O$_{4}$ (MgV$_{2}$O$_{4}$), $\arrowvert$($\varTheta$$_{CW}$)$_{\it J}$$\arrowvert$ increases from $\sim$310 K ($\sim$1145 K) to $\sim$560 K ($\sim$1550 K) as $U$ changes from 4 to 5 eV. However for CdV$_{2}$O$_{4}$, it become less than the experimentally reported value, where it changes from $\sim$40 K to $\sim$160 K as $U$ increases from 4 to 5 eV. It is well known that these spinels belong to the family of geometrically frustrated systems. Hence, the magnetic transition temperature (where the magnetic ordering occur) will be different from the Curie-Weiss ($\varTheta$$_{CW}$)$_{\it S}$ or ($\varTheta$$_{CW}$)$_{\it J}$ because of the presence degree of geometrical frustration in these spinels.               

    Now, in order to calculate the magnetic transition temperature for these vanadates, we have included the frustration by dividing the $\arrowvert$($\varTheta$$_{CW}$)$_{\it S}$$\arrowvert$ by $f$$_{\it S}$ as:
    
\begin{equation}     
(T_{N})_{\it S}=\frac{\arrowvert(\varTheta_{CW})_{\it S}\arrowvert}{f_{\it S}}
\end{equation}   
   
for $S$=1.

The plot of ($T$$_{N}$)$_{\it S}$ is shown in the Fig. 9(a) as a function of $U$. It is clear from the figure that the values of ($T$$_{N}$)$_{\it S}$ decreases continuously for all compounds as $U$ increases from 2 to 6 eV. For ZnV$_{2}$O$_{4}$ (MgV$_{2}$O$_{4}$) compounds, it decreases from $\sim$110 K ($\sim$195 K) to $\sim$96 K ($\sim$180 K) when $U$ changes from 3.5-5 eV. Similarly, it decreases from $\sim$35 K to $\sim$31 K for CdV$_{2}$O$_{4}$ as $U$ varies from 3.5 to 5 eV. From above discussion, it is also clear that the values of the ($T$$_{N}$)$_{\it S}$ for MgV$_{2}$O$_{4}$$>$ZnV$_{2}$O$_{4}$$>$CdV$_{2}$O$_{4}$ as per the experimental results. The values of ($T$$_{N}$)$_{\rm exp}$ for these compounds are also given in the Table 1. In this range of $U$, the values of ($T$$_{N}$)$_{\it S}$ are still about 2.5 (4.5) times larger than ($T$$_{N}$)$_{\rm exp}$ for ZnV$_{2}$O$_{4}$ (MgV$_{2}$O$_{4}$) compound. However for CdV$_{2}$O$_{4}$, it is close to the experimental one. Here, it is interesting to see whether the magnetic transition temperature of these compounds is improved by including the orbital degree of freedom in ($\varTheta$$_{CW}$)$_{\it S}$. Now, ($\varTheta$$_{CW}$)$_{\it S}$ has become ($\varTheta$$_{CW}$)$_{\it J}$. The values of magnetic transition temperature ($T$$_{N}$)$_{\it J}$ are now calculated by dividing $\arrowvert$($\varTheta$$_{CW}$)$_{\it J}$$\arrowvert$ by $f$$_{\it J}$ as:

\begin{equation}     
(T_{N})_{\it J}=\frac{\arrowvert(\varTheta_{CW})_{\it J}\arrowvert}{f_{\it J}}
\end{equation}

    In Fig. 9(b), we have given the plot of ($T$$_{N}$)$_{\it J}$ versus $U$. From figure, it is clear that the values of ($T$$_{N}$)$_{\it J}$ are also largest for MgV$_{2}$O$_{4}$ and smallest for CdV$_{2}$O$_{4}$ compound for whole range of $U$ studied here. Here, it is important to note that the ($T$$_{N}$)$_{\it J}$ for these compounds depend on the ratio of $\arrowvert$($\varTheta$$_{CW}$)$_{\it J}$$\arrowvert$ and $f$$_{\it J}$ and hence depends on {\it J$_{nn}$} as well as the contribution of $L$ to $J$. The values of {\it J$_{nn}$} and $L$ depends on the V-V, V-O bonds and V-O-V angles of distorted VO$_{6}$ octahedra, which are different for these compounds as discussed above in more detail. Among these spinels, lowest value of ($T$$_{N}$)$_{\it J}$ for CdV$_{2}$O$_{4}$ is expected because of large V-V, V-O bonds and V-O-V angles as compared to MgV$_{2}$O$_{4}$ and ZnV$_{2}$O$_{4}$ compounds. Also, the order of ($T$$_{N}$)$_{\it J}$ remains same as that of ($T$$_{N}$)$_{\it S}$ and ($T$$_{N}$)$_{\rm exp}$. The values of ($T$$_{N}$)$_{\it J}$ are $\sim$150 K, $\sim$60 K and $\sim$22 K for MgV$_{2}$O$_{4}$, ZnV$_{2}$O$_{4}$ and CdV$_{2}$O$_{4}$, respectively for 3.3$\leq$ $U \leq$5.3 eV. Now, it is interesting to compare the ($T$$_{N}$)$_{\it S}$, ($T$$_{N}$)$_{\it J}$ and ($T$$_{N}$)$_{\rm exp}$ for these spinels. It is evident from the Table 1 that the values of ($T$$_{N}$)$_{\rm exp}$ for MgV$_{2}$O$_{4}$, ZnV$_{2}$O$_{4}$ and CdV$_{2}$O$_{4}$ are $\sim$42 K, $\sim$40 K and $\sim$35 K, respectively. For 3.3$\leq$ $U \leq$5.3 eV, the ($T$$_{N}$)$_{\it J}$ is more closer to ($T$$_{N}$)$_{\rm exp}$ as compared to the ($T$$_{N}$)$_{\it S}$ indicates the importance of orbital degree of freedom for these compounds. From above discussion, it is also clear that the ($T$$_{N}$)$_{\it J}$ is found to be close to the ($T$$_{N}$)$_{\rm exp}$ for ZnV$_{2}$O$_{4}$ compound. However, for CdV$_{2}$O$_{4}$, ($T$$_{N}$)$_{\it J}$ is less than the ($T$$_{N}$)$_{\rm exp}$. Here, it is important to note that the calculated values of frustration indices for CdV$_{2}$O$_{4}$ compound are very small as $U$ changes from 3.3 to 5.3 eV, which indicates that the system is unfrustrated. Hence, it is not necessary to include $f$$_{\it J}$ for calculating ($T$$_{N}$)$_{\it J}$ in this system. Now, ($\varTheta$$_{CW}$)$_{\it J}$ $\approx$ ($T$$_{N}$)$_{\it J}$ and is found to be close to the experimental result.  However, only for presenting the order of ($T$$_{N}$)$_{\it J}$ for these compounds, we have included the $f$$_{\it J}$ in CdV$_{2}$O$_{4}$. For MgV$_{2}$O$_{4}$, ($T$$_{N}$)$_{\it J}$ is still about 3.5 times larger than ($T$$_{N}$)$_{\rm exp}$. The various reason for ($T$$_{N}$)$_{\it J}$$>$($T$$_{N}$)$_{\rm exp}$  for this compound are discussed below: (i) the magnetic transition temperature as well as the Curie-Weiss temperature depends on the purity of the sample. For example, ($T$$_{N}$)$_{\rm exp}$ reported by the Nishiguchi {\it et al.} is $\sim$45 K for MgV$_{2}$O$_{4}$, which is different from the Mamiya {\it et al.},\cite{Nishiguchi,Takagi,Mamiya1,Mamiya2} (ii) the values of $f$$_{\it J}$ (which are used for calculating the ($T$$_{N}$)$_{\it J}$) are calculated from the Eqns. (1) and (5), which shows the dependence of $f$$_{\it J}$ on the experimentally reported values of ($\varTheta$$_{CW}$)$_{\rm exp}$ and ($T$$_{N}$)$_{\rm exp}$. Also, the estimation of ($\varTheta$$_{CW}$)$_{\rm exp}$ depends on the linear fitting of the reciprocal magnetic susceptibility versus temperature data, which is also not consistent for this compound. For example, ($\varTheta$$_{CW}$)$_{\rm exp}$ estimated by Mamiya {\it et al.} is $\sim$600 K, which is found to less than as reported by Blasse {\it et al.}.\cite{Mamiya1,Baltzer} and (iii) the local functional (LDA+$U$) is not found to treat the magnetic interactions properly even in  MnO, where it overestimates the magnitude of these interactions as compared to experimental results.\cite{Solovyev1998} Hence, it is expected that the ($\varTheta$$_{CW}$)$_{\it J}$ is greater than the experimental one for above mentioned spinels, which may be also responsible for larger values of ($T$$_{N}$)$_{\it J}$. At last, we conclude that even having the above mentioned ambiguities, the magnetic properties of all three compounds are well explained for 3.3$\leq$ $U \leq$5.3 eV. Hence, the present study clearly shows the importance of orbital degrees of freedom in understanding the geometrically frustrated magnetic behaviour of these compounds.

\section{Conclusions}

    In conclusion, we have made an attempt to resolve the issue related to the degree of geometrical frustration present in AV$_{2}$O$_{4}$ (A $\equiv$ Zn, Cd and Mg) compounds that arises from the two experimental data (i.e. frustration indices and magnetic moments) by using detailed LDA+$U$+SOC calculations. The spin-orbit coupling was found to play an important role in deciding the magnetic properties of these compounds. The magnitude of maximum calculated values of orbital magnetic moment per formula unit for CdV$_{2}$O$_{4}$ ($\sim$1.74 $\mu$$_{B}$)$>$ZnV$_{2}$O$_{4}$ ($\sim$1.54 $\mu$$_{B}$)$>$MgV$_{2}$O$_{4}$ ($\sim$0.92 $\mu$$_{B}$). The difference between the calculated and experimentally observed magnetic moment ($\Delta$M$_{2}$=M$_{\rm total}$-M$_{\rm exp}$) was observed to be largest for MgV$_{2}$O$_{4}$ and smallest for CdV$_{2}$O$_{4}$ compound. The values of the frustration indices ($f$$_{\it J}$) by including the spin and orbital part of magnetic moment were also found to be largest for MgV$_{2}$O$_{4}$ and smallest for CdV$_{2}$O$_{4}$ compounds for $U$=3.3-5.3 eV. The consistency about the degree of geometrical frustration (which arises from $\Delta$M$_{2}$ and $f$$_{\it J}$) was obtained for this range of $U$. The order of the absolute values of nearest neighbour exchange coupling constant ({\it J$_{nn}$}) between V spins were found to be MgV$_{2}$O$_{4}$$>$ZnV$_{2}$O$_{4}$$>$CdV$_{2}$O$_{4}$, which show that the magnitude of Curie-Weiss temperature [($\varTheta$$_{CW}$)$_{\it J}$] were highest for MgV$_{2}$O$_{4}$ and smallest for CdV$_{2}$O$_{4}$ for 3.3$\leq$ $U \leq$5.3 eV. In this range of $U$, the order of magnetic transition temperature [($T$$_{N}$)$_{\it J}$] as well as the band gap were found to be in accordance with the experimental results for these compounds. At last, we conclude that all the magnetic properties studied here were found to be well explained for 3.3$\leq$ $U \leq$5.3 eV.        

\acknowledgments {S.L. is thankful to UGC, India, for financial support.}

\pagebreak

\begin{table}[ht]
\caption{Experimentally observed Curie-Weiss temperature [($\varTheta$$_{CW}$)$_{exp}$] (K), magnetic transition temperature [($T$$_{N}$)$_{exp}$] (K), frustration index (\textit{f$_{S}$}), magnetic moment per formula unit [M$_{\rm exp}$] ($\mu$$_{B}$) and calculated values of orbital magnetic moment per formula unit [M$_{\rm orbital}$] ($\mu$$_{B}$) for ACr$_{2}$O$_{4}$ (A=Zn, Mg and Cd) compounds.}
\centering 
\begin{tabular}{p{2.4cm}p{2.5cm}p{2.5cm}p{2.5cm}p{2.5cm}p{2.5cm}p{2.5cm}}
\hline
\hline
Compound&($\varTheta$$_{CW}$)$_{exp}$\cite{Takagi}&($T$$_{N}$)$_{exp}$\cite{Reehuis,Takagi}&\textit{f$_{S}$}\cite{Takagi}&M$_{\rm exp}$\cite{Wheeler}&M$_{\rm orbital}$\cite{Maitra,Pandey2011,Pandey2012}\\[0.5ex]
\hline

ZnV$_{2}$O$_{4}$&-850&40&21.3&1.26&-1.50\\
MgV$_{2}$O$_{4}$&-600&42&14.3&0.94&-0.40\\
CdV$_{2}$O$_{4}$&-400&35&11.4&2.38&-0.40\\ [1ex]
\hline

\end{tabular}
\label{table:the exp}
\end{table}

\begin{figure}
\caption{(a) Atomic arrangements of V and O atoms of AV$_{2}$O$_{4}$ (A $\equiv$ Zn, Cd and Mg) compounds in tetragonal unit cell. Each V atom is located at the center of edge sharing VO$_{6}$ octahedra (V-O bond is not connected for the sake of clarity) and (b) a schematic energy level diagram showing the splitting of $d$ levels due to crystal field and spin-orbit coupling.}
\includegraphics{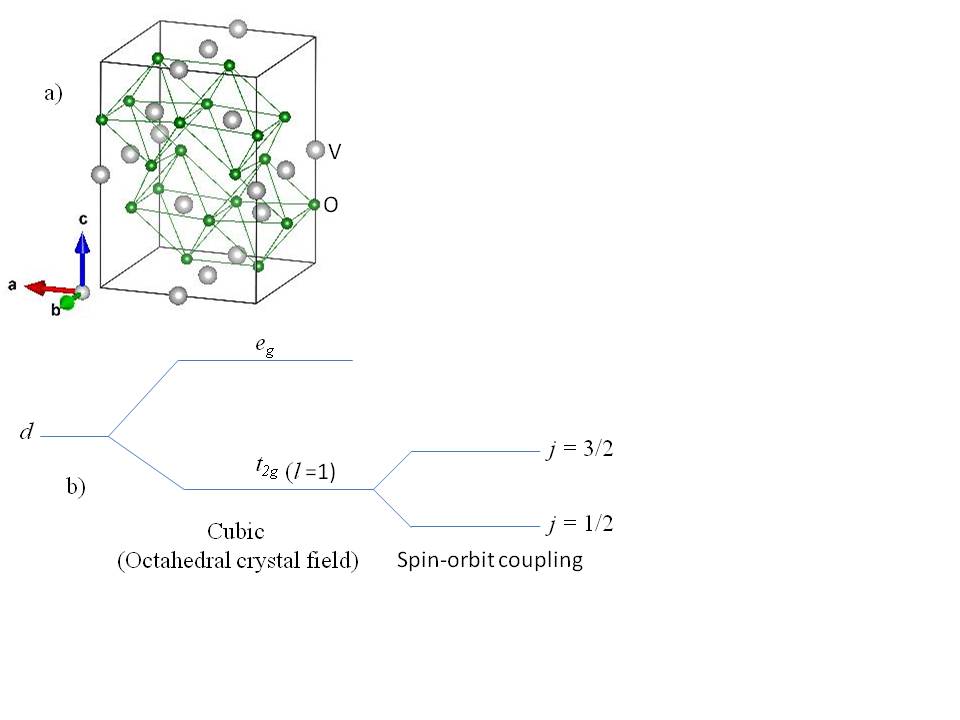}
\end{figure}

\begin{figure}
\caption{Spin (M$_{\rm spin}$), orbital (M$_{\rm orbital}$) and total (M$_{\rm total}$) magnetic moments per formula unit as a function of $U$ obtained from LDA+$U$+SOC calculations for (a) ZnV$_{2}$O$_{4}$, (b) MgV$_{2}$O$_{4}$ and (c) CdV$_{2}$O$_{4}$ compounds.}
\includegraphics{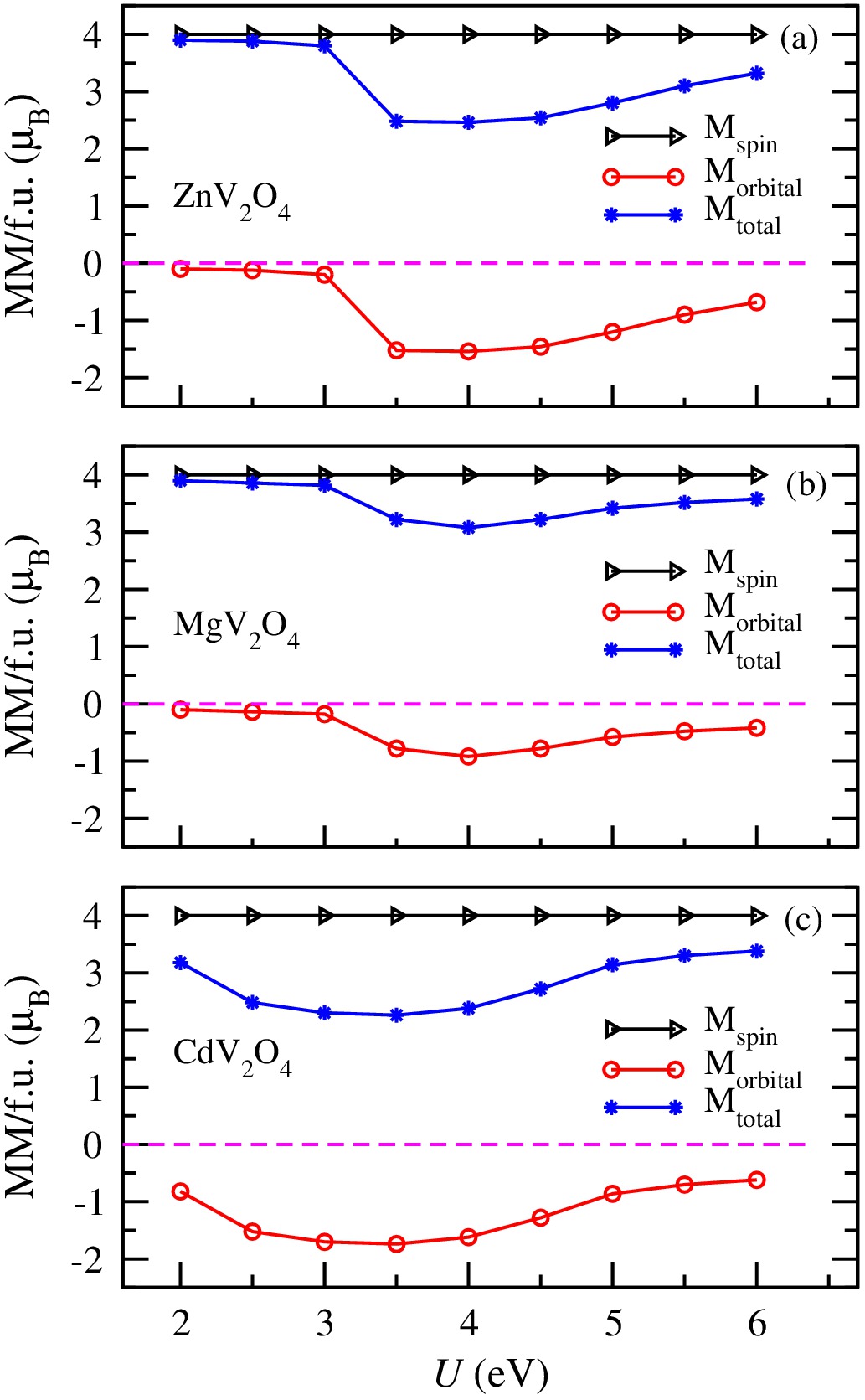}
\end{figure}

\begin{figure}
\caption{$\Delta$M$_{2}$ versus $U$ plot for AV$_{2}$O$_{4}$ (A $\equiv$ Zn, Cd and Mg) compounds, where $\Delta$M$_{2}$=M$_{\rm total}$-M$_{\rm exp}$.}
\includegraphics{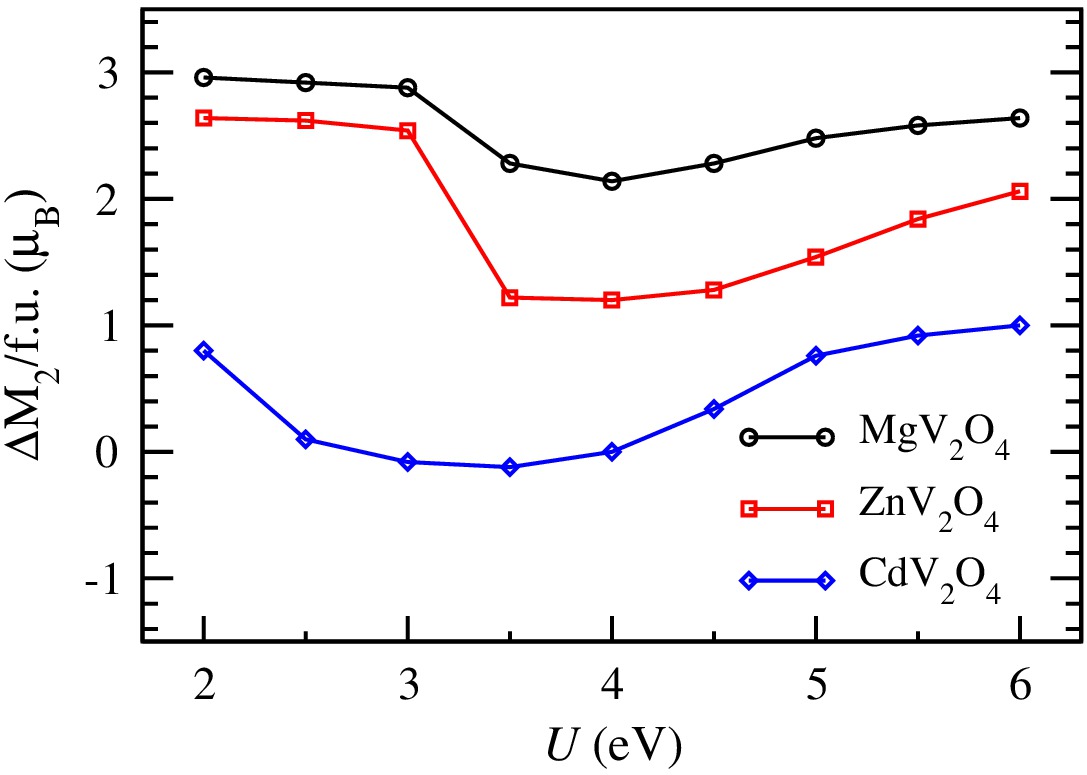}
\end{figure}

\begin{figure}
\caption{The variation of the frustration indices ($f$$_{\it J}$) with respect to $U$ for AV$_{2}$O$_{4}$ (A $\equiv$ Zn, Cd and Mg) compounds.}
\includegraphics{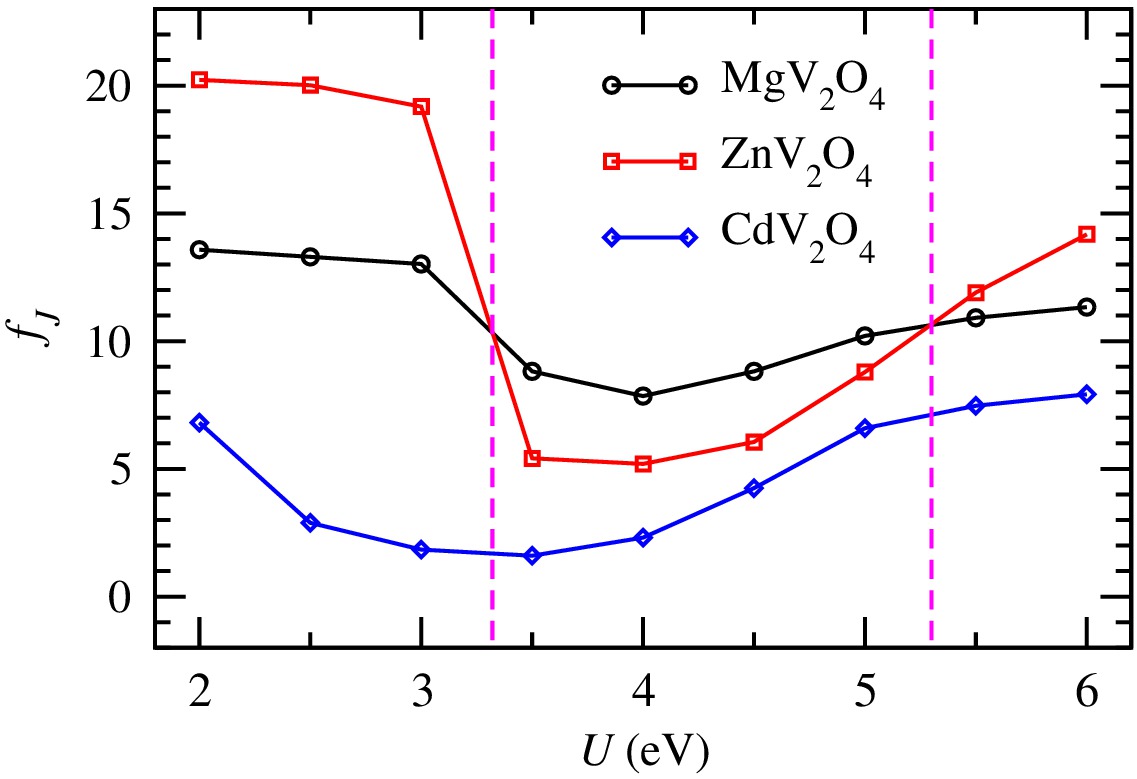}
\end{figure}

\begin{figure}
\caption{The atomic and spin arrangements of all four V atoms in the antiferromagnetic primitive unit cell (only used for calculating the nearest neighbour exchange-interaction parameter ({\it J$_{nn}$})) of AV$_{2}$O$_{4}$ (A $\equiv$ Zn, Cd and Mg) compounds.}
\includegraphics{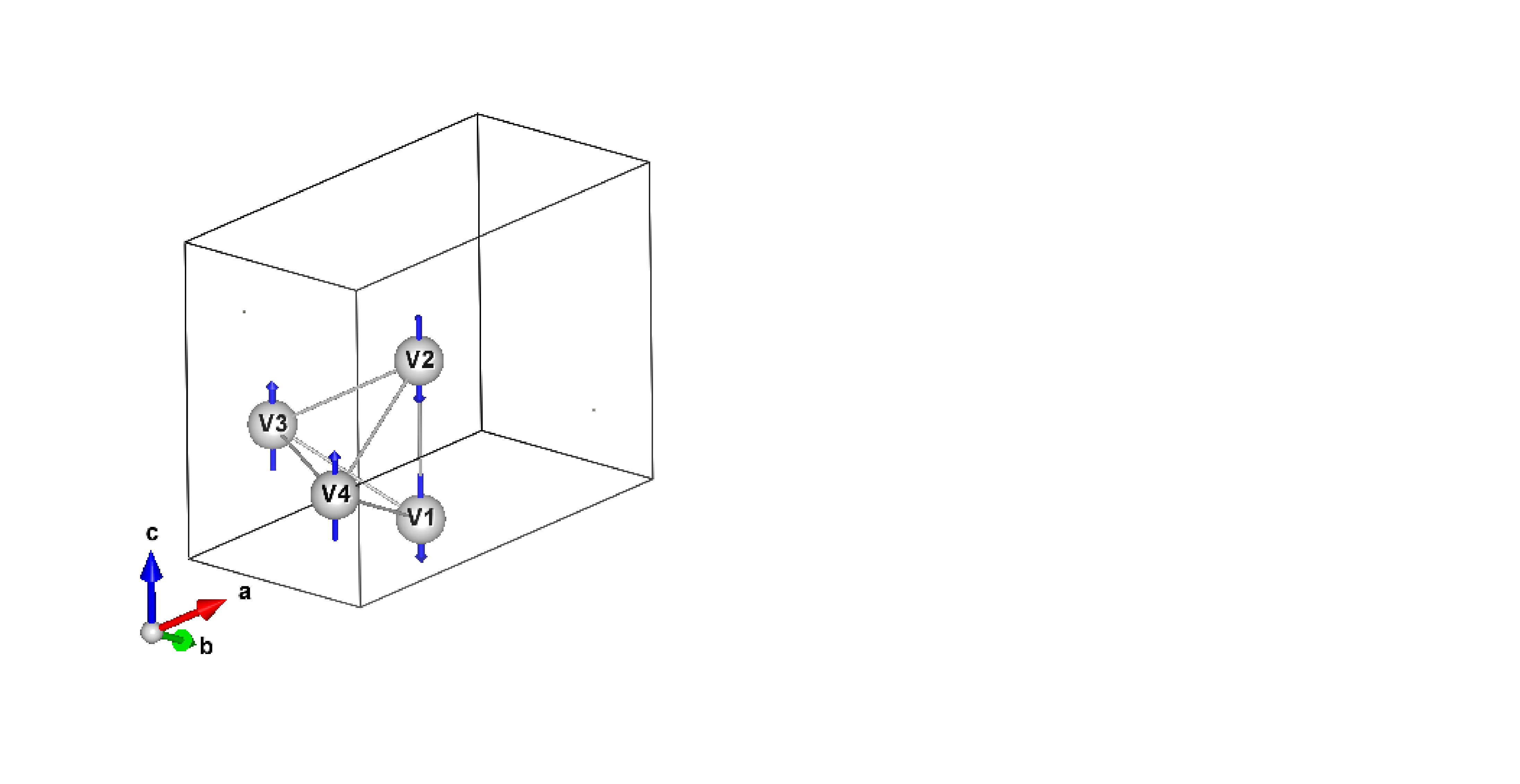}
\end{figure}

\begin{figure}
\caption{The nearest neighbour exchange-interaction parameter ({\it J$_{nn}$}) obtained by calculating the total energies per unit cell of ferromagnetic and antiferromagnetic configurations onto a classical Heisenberg model as a function of $U$ for AV$_{2}$O$_{4}$ (A $\equiv$ Zn, Cd and Mg) compounds.}
\includegraphics{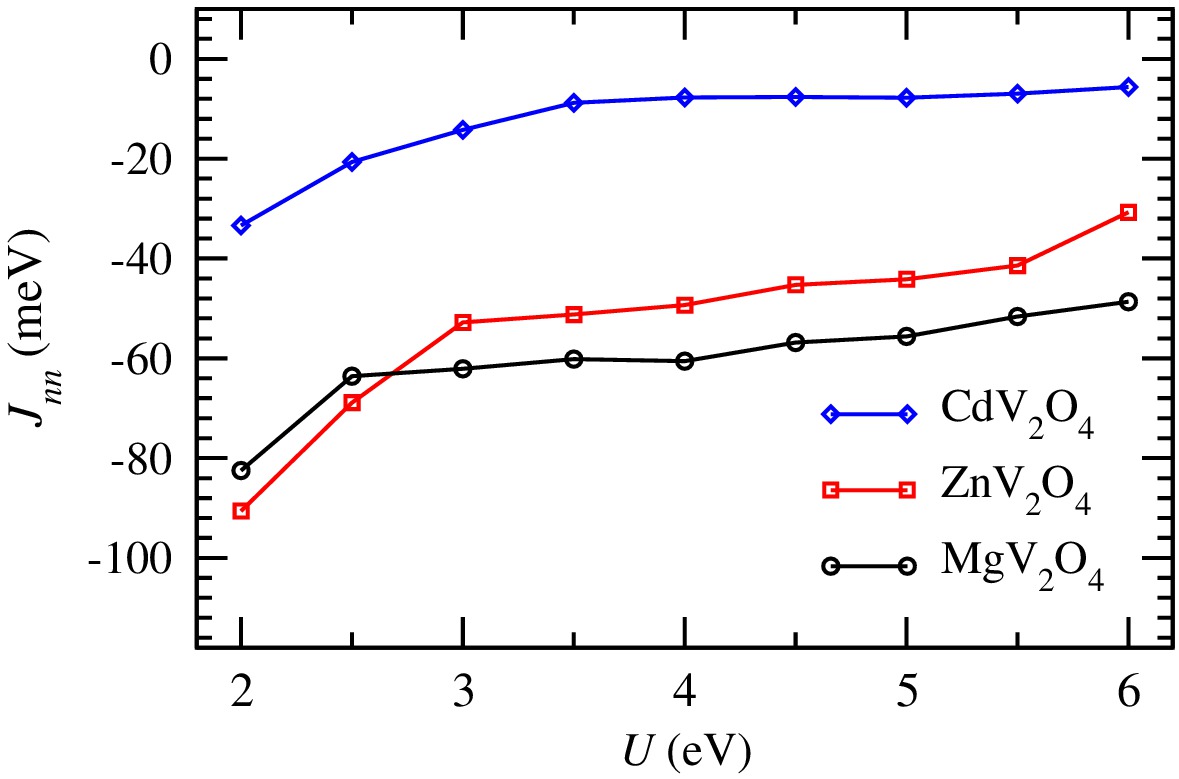}
\end{figure}

\begin{figure}
\caption{Curie-Weiss temperature [($\Theta$$_{\rm CW}$)$_{\it S}$] estimated from the calculated values of the nearest neighbour exchange-interaction parameter ({\it J$_{nn}$}) versus $U$ for AV$_{2}$O$_{4}$ (A $\equiv$ Zn, Cd and Mg) compounds.}
\includegraphics{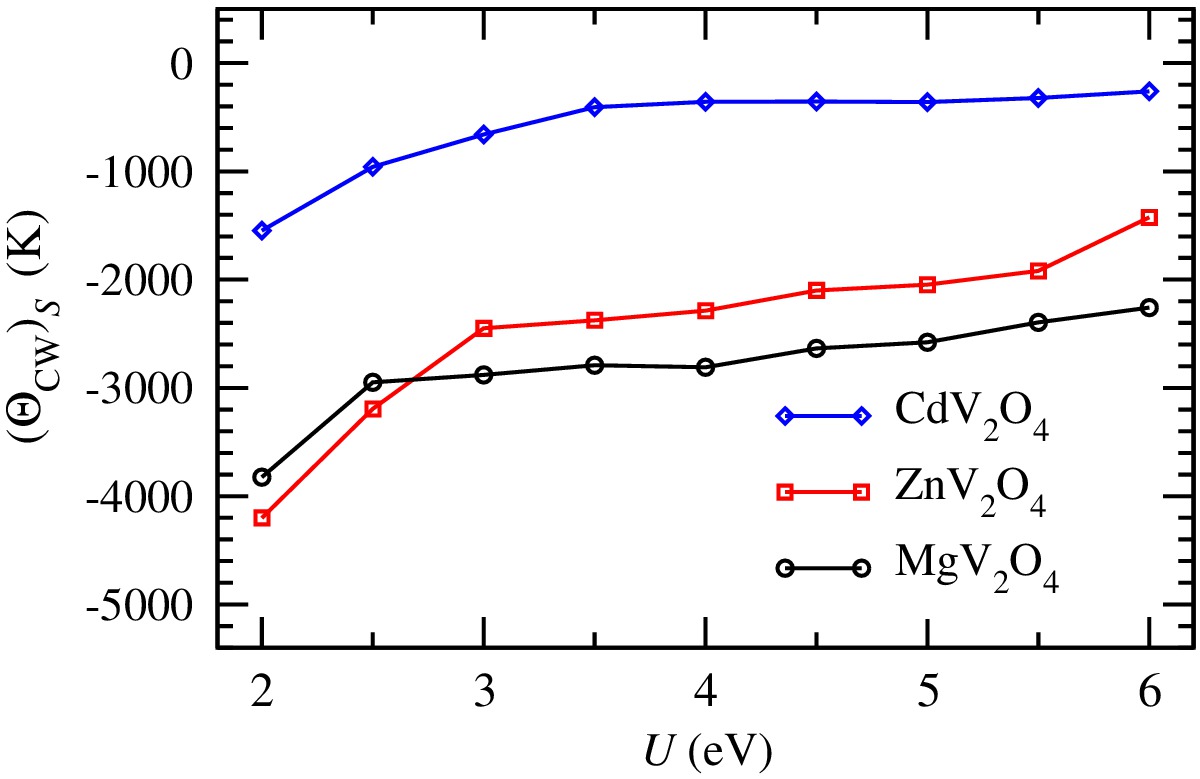}
\end{figure}
\begin{figure}
\caption{The variation of the Curie-Weiss temperature [($\Theta$$_{\rm CW}$)$_{\it J}$] as a function of $U$ for AV$_{2}$O$_{4}$ (A $\equiv$ Zn, Cd and Mg) compounds.}
\includegraphics{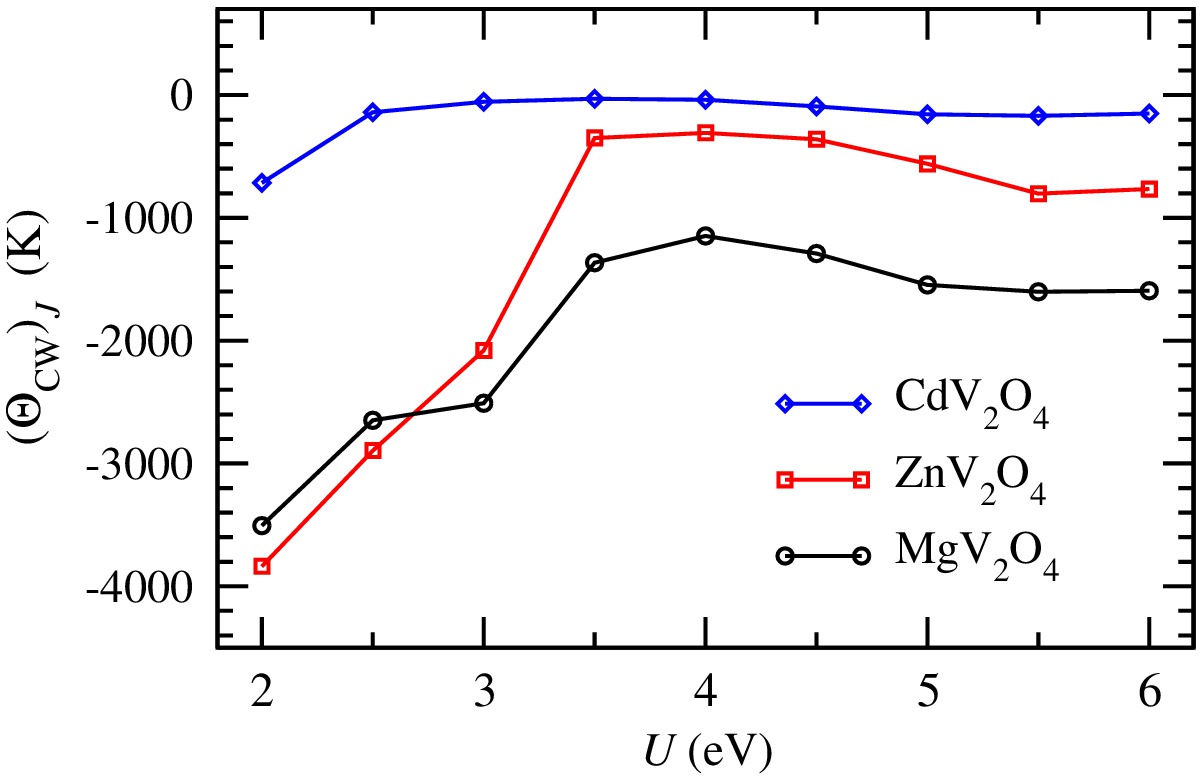}
\end{figure}

\begin{figure}
\caption{(a) The magnetic transition temperatures [($T$$_{N}$)$_{\it S}$] and (b) [($T$$_{N}$)$_{\it J}$] as a function of $U$ for AV$_{2}$O$_{4}$ (A $\equiv$ Zn, Cd and Mg) compounds.}
\includegraphics{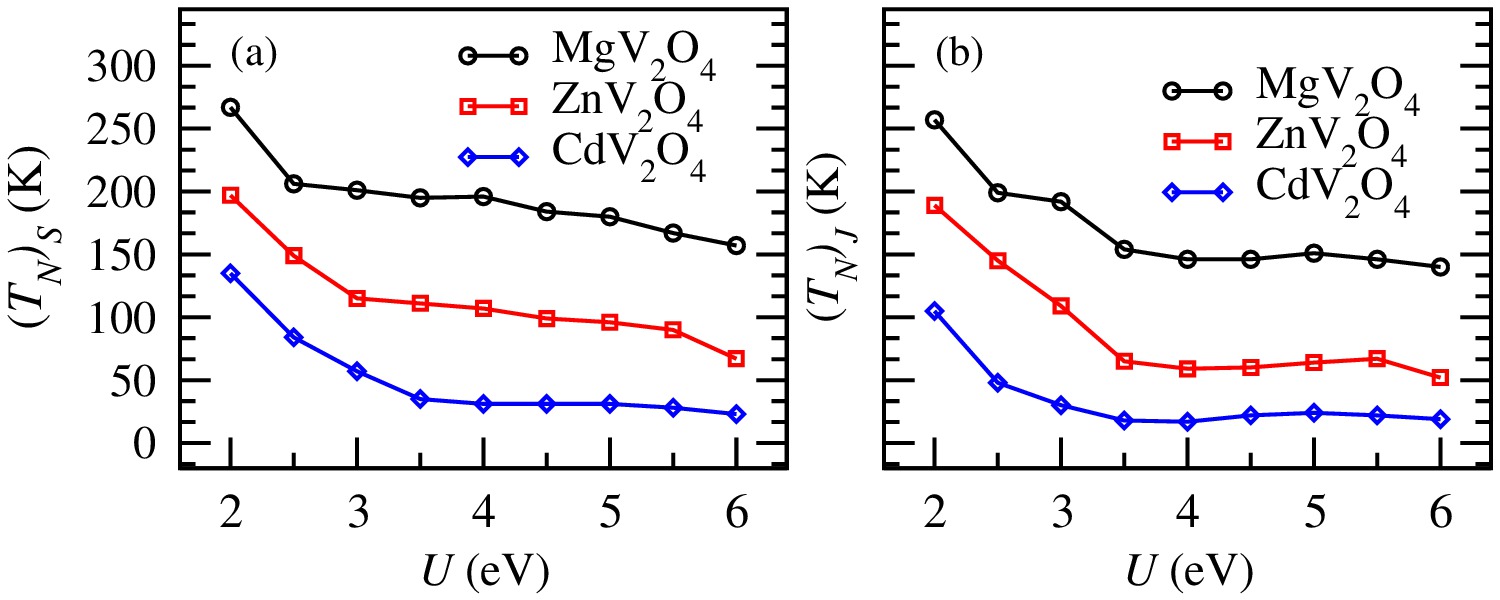}
\end{figure}

\end{document}